\documentclass[prb,twocolumn,floatfix]{revtex4}

\usepackage{amsfonts}
\usepackage{amsmath}
\usepackage{verbatim}
\usepackage[dvips]{graphicx}
\usepackage{subfigure}

\def\dd{\textrm{d}}

\begin{document}

\title{Pairing instabilities in topological insulator quantum wells}
\author{Predrag Nikoli\'c$^{1,2}$ and Zlatko Te\v{s}anovi\'{c}$^{2}$}
\affiliation{$^1$School of Physics, Astronomy and Computational Sciences,\\George Mason University, Fairfax, VA 22030, USA}
\affiliation{$^2$Institute for Quantum Matter at Johns Hopkins University, Baltimore, MD 21218, USA}
\date{\today}


\begin{abstract}

Topological insulator quantum wells with induced attractive interactions between electrons are candidate systems for the realization of novel vortex lattice states, and incompressible quantum vortex liquids with fractional excitations. These phases depend on the formation of low-energy spin-triplet Cooper pairs in the quantum well, under the combined influence of the superconducting proximity effect and Rashba spin-orbit coupling. We analyze, using the Landau-Ginzburg framework, the competition between different pairing channels stimulated by the proximity effect and argue that correlated phases of triplets are likely accessible by tuning the gate voltage. We perturbatively calculate the triplet pairing instability and show that it occurs at finite momenta. This yields a spatially inhomogeneous superconducting state, which is likely a vortex lattice of spin supercurrents. We discuss the phase diagram of topological insulator quantum wells tunable by the gate voltage.

\end{abstract}

\maketitle

\section{Introduction}

All topological insulator (TI) materials discovered so far are band-insulators\cite{Kane2005, Kane2005a, Bernevig2006, Konig2007, Hasan2010, Qi2010a, Moore2010}. They have gapless states at their boundaries\cite{Kane2005a} in common with integer quantum Hall systems, but respect the time-reversal (TR) symmetry. As in any band-insulator, their bulk excitations are conventional electron and hole quasiparticles. Since the Rashba spin-orbit coupling in TIs can be viewed as a source of an effective SU(2) magnetic field for electrons \cite{Frohlich1992, Nikolic2011, Nikolic2012}, it is natural to ask whether other observed behaviors of electrons in magnetic fields could be replicated with the TR symmetry in TIs. Especially interesting possibilities to look for are the TR-invariant incompressible quantum liquids similar to fractional quantum Hall states\cite{Levin2009, Karch2010, Cho2010, Maciejko2010, Swingle2011, Park2011, Santos2011, Neupert2011}, but with potentially novel kinds of SU(2) dynamics that could be applied in quantum computing \cite{Nikolic2012}. There are very few practical proposals and ideas for the realization of such states \cite{Pesin2010, Sun2011, Neupert2011a, Ghaemi2011}.

\begin{figure}
\centering
\includegraphics[height=1.6in]{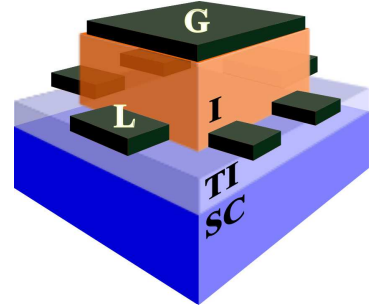}
\caption{\label{Device}The heterostructure device that could host fractional TR-invariant quantum states \cite{Nikolic2011a}. A topological insulator (TI) quantum well is sandwiched between a conventional superconductor (SC) and a conventional insulator (I). The gate (G) voltage can be used to control the state of the TI, and the topological properties of the TI can be probed via a Hall-bar setup of leads (L).}
\end{figure}

Here we investigate certain important aspects of one of these proposals, which is based on the heterostructure device shown in the Fig.\ref{Device}. This device features a TI quantum well placed in proximity to a conventional superconductor (SC). It has been suggested that such a system could be designed to host novel TR-invariant incompressible quantum liquids with fractional excitations \cite{Nikolic2011a}. The path to possible fractionalization here involves the formation of low-energy spin-triplet Cooper pairs in the TI, and the accessibility of a tunable quantum phase transition between their superconducting and insulating states. The Rashba spin-orbit coupling can turn the superconductor of spinful triplets into a vortex lattice state \cite{Nikolic2011a}. The correlated insulating states, obtained by the quantum melting of this vortex lattice, are candidates for novel fractional TIs with non-Abelian statistics \cite{Nikolic2011a, Nikolic2012}.

A rather simplified picture of triplet superconductivity in the TI quantum well is the following. Phonons of the proximate SC material mediate weak attractive interactions between the TI's electrons. This can in principle create Cooper pairs in various channels inside the TI. Singlet pairs are favored in typical situations, but the TI quantum well gives triplets a special boost. First of all, the triplets in question are formed by two nearby electrons that come from the opposite surfaces of the quantum well. The electrons' surface degree of freedom (not a quantum number) allows the existence of ``inter-orbital'' triplets whose size is comparable to that of singlets without violating the Pauli exclusion principle. Second, and more important, the spinful triplets experience the Rashba spin-orbit coupling in the TI. Consequently, their energy can be lowered in proportion to the momentum they carry, as long as their spin has the proper orientation. The only fundamental limit to such an energy gain is set by the momentum cut-off related to the crystalline structure of the material. Certain triplet modes can be made to condense, before or after singlets abundantly condense. The gate voltage applied in the device from Fig.\ref{Device} can be used to control the chemical potential in the TI, and thus drive the condensation of triplets.

The purpose of this paper is to verify the above picture of triplet superconductivity by a simple model calculation, and thus support the proposed route to fractional TIs of Cooper pairs \cite{Nikolic2011a}. We will employ field theory methods to argue that gated TI quantum wells can be used to condense triplets. Then we will focus on the triplet pairing instability and elucidate its character. One of our main results is that triplets condense at \emph{finite momenta} into a superconducting state that carries ``helical'' spin-currents and naturally leads to the formation of a vortex lattice in equilibrium. Our present analysis is phenomenological and provides a proof of principle rather than a microscopic evidence of feasibility in realistic heterostructures. The latter has been discussed in the Ref.\cite{Nikolic2011a}, together with the mechanism for obtaining fractionalized quantum liquids. The prospects for fractional TIs are only a motivation for the present study.

The physics pursued here should be distinguished from that of ``topological superconductors'' which can harbor Majorana quasiparticles. The latter can be obtained by the superconducting proximity effect on the surface of a \emph{bulk} TI \cite{Fu2008, Stanescu2010a, Sau2010c, Linder2010, Sato2010, Beenakker2012, Kasumov1996, Koren2011, Qu2011, Sacepe2011, Yang2011, Zhang2011, Veldhorst2012, Wang2012, Williams2012}. In contrast, we consider TI \emph{quantum wells}, where the surface hybridization ensures a bandgap in the TI's electronic spectrum \cite{Zhang2010, Zhang2009c, Kong2010, Hong2010, Liu2011a, Cho2011a}. The induced attractive interactions among the TI's \emph{insulating} electrons can stabilize correlated superconducting as well as \emph{insulating} states of low-energy triplet Cooper pairs \cite{Nikolic2010, Nikolic2010b}. The phase transition between these states \emph{can be driven} by the gate voltage. This would allow experimental access to the anticipated fractional insulators.

The expected triplet superconductivity in TI quantum wells may have certain similarities with the superfluid $^3$He-$B$, despite different dimensionality. Our system could naturally host abundant topological defects of spin supercurrents, which have been studied both theoretically \cite{Mineev1986, Mineev1992, Salomaa1987} and experimentally \cite{Kondo1992} in $^3$He-$B$.

We start our analysis in the section \ref{secModel} with an outline of the rich proximity effect physics in TI quantum wells. This will provide a justification for the model we use in calculations. The following section \ref{secCompetition} considers the competition between singlet and triplet pairing in the TI. There we argue, using the Landau-Ginzburg approach, that triplet superconductivity can be obtained as a matter of principle in TI quantum well heterostructures, with or without a coexisting singlet superconductivity. The technical section \ref{secTriplet} shows by perturbation theory that the triplet pairing instability occurs at finite momenta. Readers not interested in detailed calculations can skip this section and go directly to the section \ref{secPhDiag} that discusses the phase diagram of triplets. There we explain why the triplet condensation likely yields a TR-invariant vortex lattice state, and mention its connection to the anticipated fractional TI states. Finally, we discuss some limitations of our approach and summarize conclusions in the section \ref{secDiscussion}.

\bigskip

\section{Proximity effect in topological insulator quantum wells}\label{secModel}

An isolated TI quantum well made from a material such as Bi$_2$Se$_3$ or Bi$_2$Te$_3$ can be modeled by the second-quantized Hamiltonian
\begin{equation}\label{SO}
H_0 = \psi^\dagger \Bigl\lbrack v\,\hat{{\bf z}}({\bf S}\times{\bf p})\tau^{z}+\Delta\tau^{x}-\mu \Bigr\rbrack \psi \ ,
\end{equation}
where ${\bf S} = \frac{1}{2}\boldsymbol{\sigma}$ is the spin operator ($\hbar=1$). The Pauli matrices $\sigma^a$ and $\tau^a$ ($a \in \lbrace x,y,z \rbrace$) act in the spin and ``orbital'' spaces respectively. The spinor field operator $\psi$ has four components $\psi_{\tau\sigma}$ that correspond to the TI's low-energy degrees of freedom: two spin states labeled by the eigenvalue $\sigma = \pm 1$ of $\sigma^z$, and two ``orbital'' states labeled by the eigenvalue $\tau = \pm 1$ of $\tau^z$. The orbital index is equivalent to the top/bottom surface of the quantum well. This minimal model derives its dynamics only from the Rashba spin-orbit coupling of strength $v$. A more realistic model includes kinetic energy, and can be conveniently written as \cite{Nikolic2011a}
\begin{equation}\label{SO2}
H_0' = \psi^\dagger\left\lbrack\frac{({\bf p}-\tau^z\boldsymbol{\mathcal{A}})^2}{2m}+\Delta\tau^{x}-\mu\right\rbrack\psi \ ,
\end{equation}
using a static external SU(2) gauge field
\begin{equation}\label{GaugeField}
\boldsymbol{\mathcal{A}} = -mv (\hat{\bf z} \times {\bf S})
\end{equation}
to represent the Rashba spin-orbit coupling. Note that the electron's orbital index $\tau^z$ plays the role of an SU(2) charge in this gauge-theory representation. The two TI's surfaces are close enough (a few quintuple layers apart) to hybridize via the tunneling term $\Delta \neq 0$. The chemical potential $\mu$ inside the quantum well can be tuned by applying the gate voltage in the device from the Fig.\ref{Device}. The TI quantum well is a band-insulator if $\mu$ lies within its bandgap created by the surface hybridization $\Delta$. Note that $H_0$ applied to a thick TI slab ($\Delta = 0$) would capture its topologically protected surface states, but the tunability of $\mu$ by a gate voltage (and the physics we discuss below) would be lost.

The dynamics of electrons in the TI quantum well is modified by their proximity to the SC material in the Fig.\ref{Device}. First of all, the SC material presents explicit U(1) symmetry breaking to the TI's electrons. This results in the conventional proximity effect, the leakage of the singlet superconducting order parameter from the SC into the TI. The effective action of the TI acquires pairing terms $\Delta_{\textrm{s},\tau\tau'}^* \psi_{\tau\uparrow} \psi_{\tau'\downarrow} + h.c.$ that explicitly break the U(1) symmetry. Secondly, \emph{two-body interactions} among the TI's electrons become renormalized. Any charge-carrying excitation in the quantum well can displace the SC's atoms near the interface, and thus attractively interact with another TI's excitation via emission and absorption of the SC's phonons. The Coulomb origin of this interaction makes it spin-independent, and its range is short due to screening that takes place in the SC. Virtual Cooper pair tunneling from the SC into the TI dynamically generates additional short-range pairing forces in the TI. A careful study of all these interactions shows that their strength can be sufficient to overcome Coulomb forces and cause pairing in the TI at experimentally accessible temperatures, independently of the conventional proximity effect \cite{Nikolic2012b}. This only requires choosing a superconductor for the heterostructure design whose critical temperature is sufficiently large (such as MgB$_2$ or perhaps even Nb).

The conventional proximity effect is very significant at a SC interface with a \emph{bulk} three-dimensional TI. The TI's surface in that case has topologically protected modes, which are gapless and thus strongly affected by the explicit pairing energy scale $\Delta_{\textrm{s}}$. This yields interesting surface superconductivity capable of hosting Majorana modes bound to topological defects \cite{Fu2008}. However, the analogous effect can be very weak in our TI \emph{quantum well} system. Typical surface hybridization gaps can be of the order of $\Delta \sim 100 \textrm{ meV}$, which is much larger than the typical pairing scale $\Delta_{\textrm{s}} \sim 1 \textrm{ meV}$ in the best conventional superconductors. Therefore, the explicit U(1) symmetry breaking is only a very weak perturbation that by itself produces a negligible order parameter in the TI if the chemical potential sits deep within the bandgap.

Significant pairing in the TI quantum well can occur only when the chemical potential is brought close to the conduction or valence band of its electronic spectrum. However, this makes interaction effects important as well. Without having a faithful microscopic model of the heterostructure, we cannot sensibly compare the relative strengths of the conventional proximity effect and induced interactions. Hence, we must keep our analysis qualitative and very general. The best way to accomplish that goal is to represent all of the above effects in the Landau-Ginzburg theory of the TI quantum well, and then explore different possibilities for the ground state. Using this approach, we will find that the generated attractive interactions have a significant impact on the triplet pairing sector.

Being attractive and largely spin-independent, the induced interactions among the TI's electrons can in principle create Cooper pairs in various channels. Singlet channels are also helped by the conventional proximity effect. The available TI's electronic degrees of freedom give rise to the following low-energy pairing channels (represented by Cooper pair field operators):
\begin{itemize}
  \item intra-orbital singlets: $\phi_\tau = \frac{1}{\sqrt{2}} \epsilon_{\alpha\beta} \psi_{\tau\alpha} \psi_{\tau\beta}$ \ ,
  \item inter-orbital singlets: $\phi_0 = \frac{1}{\sqrt{2}} \epsilon_{\alpha\beta} \psi_{+\alpha} \psi_{-\beta}$ \ ,
  \item ``spinful'' triplets: $\eta_\sigma = \psi_{+\sigma} \psi_{-\sigma}$ \ ,
  \item symmetric triplets: $\eta_0 = \frac{1}{\sqrt{2}} (\psi_{+\uparrow} \psi_{-\downarrow} + \psi_{+\downarrow} \psi_{-\uparrow})$ \ .
\end{itemize}
In all cases, the coordinate dependence of the paired electron wavefunction is $s$-wave-like, allowing the two electrons to overlap and take advantage of the short-range interactions. The Pauli exclusion principle is satisfied either by the electrons' spin ($\sigma$) or ``orbital'' ($\tau$) degree of freedom. There are three kinds of singlet pairs $\phi_-, \phi_0, \phi_+$, and three kinds of inter-orbital triplet pairs $\eta_-$, $\eta_0$, $\eta_+$. The states $\eta_m$ correspond to the three possible values $m\in\lbrace -1,0,1 \rbrace$ of the triplet $S^z$ spin projection. Note that these channels are not the same as the paired eigenmodes that could exist in our system.

A pairing channel is important for the low-energy dynamics if it admits long-lived gapped bosonic excitations, or becomes condensed. All such channels should be included in the Landau-Ginzburg theory of the TI quantum well. Fermionic excitations can also be included in the theory, but this is unnecessary if they linger only at high energies. Our specific interest here are the superconducting phase transitions out of the TI's insulating state. We will drive such transitions by manipulating the chemical potential $\mu$ that initially sits deep inside the TI's bandgap. Even as $\mu$ is tuned toward an electron band via the gate voltage, fermionic excitations remain gapped all the way until the superconducting transition. This will justify our initial treatment of electrons as high-energy excitations, and the ensuing Landau-Ginzburg description of the low-energy Cooper pair dynamics.

Bosonic Cooper pair excitations are made possible at low energies by the two-dimensional geometry of the TI quantum well. It is known that any amount of short-range attractive interactions between two isolated non-relativistic particles always produces a bound state in two dimensions \cite{L1977}. Therefore, the TI's Cooper pairs can be genuine two-body bound states (the analogous bound states are not possible in three-dimensional superconductors for realistic interaction strengths). Bringing the chemical potential about a binding energy away from the electronic conduction or valence band will generally condense these Cooper pairs. Fermions remain gapped ``high-energy'' excitations across this transition. Note that the binding energy cannot be larger than the SC's pairing gap. The TI's Cooper pairs, therefore, live at energies within the SC's gap and their two-dimensional dynamics near the transition is not jeopardized by the presence of higher energy extended states in the SC.

The character of the lowest-energy Cooper pairs that condense first depends on their spin. Singlets have the lowest energy at zero momentum, and become unstable to decay beyond some finite momentum when their energy per electron penetrates a fermionic band. Triplets are more complicated. We will find that one triplet mode becomes enhanced by the Rashba spin-orbit coupling and exhibits an inverted behavior, having the lowest energy at largest momenta. By its nature, this triplet mode is very competitive with singlets.

\section{Singlet vs. triplet superconductivity inside the TI}\label{secCompetition}

This section explores the competition between singlets and triplets. Our goal is to argue that triplet superconductivity is very likely accessible in the device of Fig.\ref{Device} merely by tuning the gate voltage at low temperatures. To that end, let us consider the general imaginary-time Landau-Ginzburg action of stable low-energy Cooper pairs in the regime where they dominate the dynamics. This action describes only the TI quantum well. It can be readily constructed from the symmetry considerations alone, without knowing the particular values of its various coupling constants \cite{Nikolic2011a}:
\begin{eqnarray}\label{LG2}
S_{\textrm{eff}} \!\! &=& \!\! \int \dd\widetilde{t} \; \dd^2 r \; \Bigl\lbrace
    \phi^* \partial_0 \phi +
    K_{\textrm{s}} \left\vert\boldsymbol{\nabla}\phi\right\vert^2
  + t_{\textrm{s}} |\phi|^2
\\ \!\! &-& \!\! \Delta_{\textrm{s}}^*\phi -\Delta_{\textrm{s}}^{\phantom{*}}\phi^* \nonumber
\\ \!\! &+& \!\! \eta^\dagger \partial_0 \eta
   +K_{\textrm{t}}\Bigl\lbrack\left(\boldsymbol{\nabla}-i\boldsymbol{\mathcal{A}}\right)\eta\Bigr\rbrack^{\dagger}
       \Bigl\lbrack\left(\boldsymbol{\nabla}-i\boldsymbol{\mathcal{A}}\right)\eta\Bigr\rbrack
   +t_{\textrm{t}}\eta^{\dagger}\eta \nonumber \\
\!\! &+& \!\! U_{\textrm{s}} |\phi|^4 + U_{\textrm{t}}(\eta^{\dagger}\eta)^{2}
   +t_{\textrm{s}}' |\phi|^2 \eta^{\dagger}\eta \Bigr\rbrace + S_{\textrm{nm}} + S_{\textrm{p}} \nonumber
\end{eqnarray}
The idealized minimal TI model (\ref{SO2}) has a formal spin SU(2) gauge symmetry, which is merely a mathematical tool to express the Rashba spin-orbit coupling. Nevertheless, if no perturbation violated this symmetry ($S_{\textrm{p}}=0$), it would have to survive in the above Landau-Ginzburg action as written. Triplet Cooper pairs carry spin, so they must couple to the static external gauge field (\ref{GaugeField}) expressed in the spin $S=1$ representation. The SU(2) generators $S^a$ involved in (\ref{GaugeField}) are to be interpreted now as the $3\times 3$ matrices of spin projections:
\begin{eqnarray}\label{SpinOperators}
&&
  S^x = \frac{1}{\sqrt{2}}\left(\begin{array}{ccc}0&1&0\\1&0&1\\0&1&0\end{array}\right) \quad,\quad
  S^y = \frac{1}{\sqrt{2}}\left(\begin{array}{ccc}0&-i&0\\i&0&-i\\0&i&0\end{array}\right) \nonumber \\
&& ~~~~~~~~~~~~~~~~~~~
  S^z = \left(\begin{array}{ccc}1&0&0\\0&0&0\\0&0&-1\end{array}\right) \ .
\end{eqnarray}
instead of the $2\times 2$ Pauli matrices halved. We will rediscover this by a concrete calculation in the section \ref{secTriplet}. In plain language, triplets experience the Rashba spin-orbit coupling. The $S_{\textrm{nm}}$ part of the action contains non-minimal TR-invariant couplings between the matter and gauge fields:
\begin{eqnarray}\label{LG2b}
S_{\textrm{nm}} \!\!&=&\!\! \int \dd\widetilde{t} \, \dd^2 r \; \Bigl\lbrack
  \left( a_1 + a_2|\phi|^2 \right) \eta^{\dagger}\Phi_{0}^{2}\eta \\
&& ~~~ +b_{1}(\eta^{\dagger}\Phi_{0}\eta)^{2}+b_{2}(\eta^{\dagger}\Phi_{0}^{2}\eta)^{2} \Bigr\rbrack \ . \nonumber
\end{eqnarray}
Here, $\Phi_0$ is the SU(2) ``magnetic'' flux matrix of the gauge field (\ref{GaugeField}):
\begin{equation}\label{Flux}
\Phi_0 = \epsilon^{0\mu\nu} \Bigl( \partial_\mu \mathcal{A}_\nu - i \mathcal{A}_\mu \mathcal{A}_\nu \Bigr) \propto S^z \ .
\end{equation}
All perturbations that violate the SU(2) symmetry are collected in the $S_{\textrm{p}}$ part of the action and will be neglected in our formal analysis. We will discuss important aspects of their effect in the section \ref{secDiscussion}.

For simplicity, we keep track of only the lowest-energy singlet mode $\phi$ in (\ref{LG2}), which is a certain linear combination of the three singlet channels introduced in the previous section. Triplet modes are combined into the spinor $\eta = (\eta_+, \eta_0, \eta_-)$. The ``stiffness'' couplings $K_{\textrm{s}}$, $K_{\textrm{t}}$, gap scales $t_{\textrm{s}}$, $t_{\textrm{t}}$, and interactions $U_{\textrm{s}}$, $U_{\textrm{t}}$, $t_{\textrm{s}}'$ are phenomenological parameters. The conventional superconducting proximity effect allows the charge U(1) symmetry violation via $\Delta_{\textrm{s}} \neq 0$.

Let us now imagine that the TI quantum well is initially in its band-insulating state. We will gradually change the gate voltage to drive the TI into a superconducting state. A fraction $V_{\textrm{g}}$ of the gate voltage, determined by the heterostructure geometry, will be applied across the quantum well. It directly controls all chemical potential couplings, because all fields carry the charge $2e$:
\begin{equation}
t_{\textrm{s}} = t_{\textrm{s}0}-2eV_{\textrm{g}} \quad,\quad t_{\textrm{t}} = t_{\textrm{t}0}-2eV_{\textrm{g}} \ .
\end{equation}
Since we do not know the values of $t_{\textrm{s}0}$ and $t_{\textrm{t}0}$, we cannot a priori tell whether singlets or triplets will condense first as we manipulate the gate voltage.

Triplets are typically inferior to singlets in condensed matter systems. However, the spin-orbit coupling in TIs gives them a special advantage. It creates two helical triplet modes whose spin is perpendicular to their momentum, analogous to the Dirac electron modes in band-insulating TIs. Since the Rashba spin-orbit coupling acts as a momentum-dependent ``Zeeman effect'', one helical mode has energy that decreases with its momentum $q$:
\begin{equation}\label{Ehel}
E_{\textrm{t}}(q) = t_{\textrm{t}} + \Delta t_{\textrm{t}} - v'q + \mathcal{O}(q^2) \ ,
\end{equation}
where $v' = 2mvK_{\textrm{t}}$, and $\Delta t_{\textrm{t}} =  \frac{1}{2} K_{\textrm{t}} mv^2$ arises from the ``diamagnetic'' term $K_{\textrm{t}} \eta^\dagger \boldsymbol{\mathcal{A}}^2 \eta$ in (\ref{LG2}). We will derive (\ref{Ehel}) by an explicit calculation in the section \ref{secTriplet}, which employs a more microscopic model to determine some relationships between various couplings in (\ref{LG2}). A helical triplet condensate forms at momenta $q \ge q_{\textrm{min}}$ where $E_{\textrm{t}}(q) \le 0$, provided that such momenta lie below the cut-off momentum $\Lambda$ set by the crystalline lattice. The other helical mode's energy $E_{\textrm{t}}'(q) = t_{\textrm{t}} + \Delta t_{\textrm{t}} + v'q$ increases with momentum, so it belongs to the uninteresting high energy spectrum. Even a naturally large triplet gap scale $t_{\textrm{t}}$ is not likely to jeopardize the helical condensate because $\Lambda$ is large. In contrast, singlets can condense only at the zero momentum provided that their gap $t_{\textrm{s}}$ is closed by tuning the gate voltage. We will neglect here the non-tunable and unavoidable singlet condensation caused by the direct order parameter coupling $\Delta_{\textrm{s}}$; it is made small by the TI's hybridization bandgap.

We are ultimately interested in triplets, but let us consider the worst case scenario. Suppose that singlets condense first from an insulating state of the TI, at some gate voltage that makes $t_{\textrm{s}} < 0 < E_{\textrm{t}}(q)$ for all $q<\Lambda$. The singlet order parameter can be estimated from the mean-field approximation
\begin{equation}
|\phi|^2 = \frac{|t_{\textrm{s}}|}{2U_{\textrm{s}}} \ .
\end{equation}
A further increase of the gate voltage only increases $|\phi|^2$, and tends to screen out the triplets via the singlet-triplet repulsive interaction $t_{\textrm{s}}'>0$. The actual energy of helical triplets is:
\begin{eqnarray}
E_{\textrm{t}}^{\textrm{eff}}(q) \!\!&=&\!\! t_{\textrm{t}0}+\Delta t_{\textrm{t}}-2eV_{\textrm{g}}+t_{\textrm{s}}'|\phi|^2-v'q \\
  \!\!&=&\!\! t_{\textrm{t}0}+\Delta t_{\textrm{t}}-2eV_{\textrm{g}}
             +\frac{t_{\textrm{s}}'}{2U_{\textrm{s}}}|t_{\textrm{s}0}-2eV_{\textrm{g}}|-v'q \ . \nonumber
\end{eqnarray}
If $t_{\textrm{s}}'/2U_{\textrm{s}}<1$, then a further increase of the gate voltage will eventually make $E_{\textrm{t}}^{\textrm{eff}}(q)<0$ at large $q$ and hence condense the triplets. This condition is likely satisfied in realistic systems owing to the Pauli exclusion among the electron constituents of Cooper pairs. Two singlet pairs have their electrons in exactly the same quantum states and hence repel each other relatively strongly at short distances ($U_{\textrm{s}}$), when their wavefunctions begin to overlap. The repulsion $t_{\textrm{s}}'$ between a singlet and a triplet pair is weaker than $U_{\textrm{s}}$, because triplet's electrons are in different quantum states than the singlet's electrons, and cause less frustration via the Pauli exclusion in wavefunction overlaps.

Therefore, triplets can be likely condensed even in the presence of a pre-formed singlet condensate in the TI. Note that the electrons needed for the new triplet Cooper pairs are simply supplied by the nearby SC material. Similarly, if triplets were to condense first, singlets could condense later at a larger $V_{\textrm{g}}$. Coulomb interactions are not an obstacle for the simultaneous singlet and triplet condensation because the gate cannot be efficiently screened in a very thin quantum well.

Given that some triplets get an energy boost from the Rashba spin-orbit coupling, which we will analyze in the section \ref{secTriplet}, it is not inconceivable that they could condense first, out of the TI's band-insulating instead of the singlet-superconducting state (we are neglecting $\Delta_{\textrm{s}}$). The formal appearance of (\ref{Ehel}) should not be taken as a sign that triplets might have a too large energy $t_{\textrm{t}0}+\Delta t_{\textrm{t}}$ at zero momentum to ever condense. The equation (\ref{Ehel}) is appropriate for the particular formulation of the Landau-Ginzburg theory (\ref{LG2}), but the net value of $t_{\textrm{t}0}+\Delta t_{\textrm{t}}$ is determined by very similar microscopic processes as the analogous ``chemical potential'' $t_{\textrm{s}}$ of singlets. The condensation of triplets in the absence of a singlet condensate would be ideal for the purposes of their experimental detection because some aspects of their charge dynamics could be measured, such as conductivities. If instead singlets condensed first, they would mask the charge dynamics of triplets. Nevertheless, singlets cannot screen spin, so the non-trivial spin dynamics of triplets is still measurable.

\section{The triplet pairing instability}\label{secTriplet}

The following discussion focuses on the dynamics of triplet Cooper pairs. We will assume for simplicity that singlets are not condensed and explore the mechanism and conditions for triplet condensation. We will perturbatively calculate the zero-temperature triplet pairing susceptibility in a band-insulating TI quantum well, treating the proximity-induced attractive interactions, quantum well band-gap and gate voltage as tunable parameters. We will show that there are realistic circumstances in which triplets undergo an unconventional superconducting transition by condensing at finite momenta, while fermionic excitations remain gapped.

The strategy is to calculate the Cooper pair Green's function in a band-insulating state, and determine from it the poles of the spin-orbit-coupled triplet modes. If real poles (mode energies) exist well below the fermion excitation gap, they dominate the low-energy dynamics. We will indeed discover condensation instabilities in which a real pole crosses zero energy at some momentum, while the fermion gap remains finite and keeps quasiparticles in the high-energy spectrum. If this instability occurred at zero momentum, the ensuing second-order transitions would belong to the bosonic mean-field or XY universality class depending on how the parameters are tuned \cite{Nikolic2010, Nikolic2010b}. However, the finite-momentum instability is more complicated. Our perturbation theory will still see the transition as a second order one, but we will argue that certain non-perturbative effects are important and likely turn the transition into a first-order one. The expected triplet superconducting state is actually a vortex lattice rather than a uniform condensate.

Since our working assumption is that the ground-state is a band-insulator, all modes are gapped and the role of their quantum fluctuations is merely to renormalize the values of various parameters in the theory, such as the characteristic spin-orbit velocity $v$, electron band-gap $\Delta$, chemical potential $\mu$, and the short-range interaction couplings. No qualitative changes of the coherent parts of various propagators can occur until a phase transition is encountered. Consequently, we can use a simple one-loop perturbative calculation to determine the qualitative dependence of the Cooper pair propagator on the momentum, frequency, and the renormalized parameters. This is enough to reveal if or when the bosonic modes exist below the fermion excitation gap. The effects of (perturbative) fluctuations are qualitatively included at all orders of perturbation theory because we use the full renormalized propagators and vertices in calculations, neglecting only their incoherent parts.

Even though we analyze the triplet condensation instability of a band-insulating state, we note that our main conclusions are qualitatively valid for the analogous triplet instability in the presence of a pre-formed singlet condensate. In the latter case, we would only have to replace gapped electron excitations with gapped Bogoliubov quasiparticles of the singlet superconductor in calculations. The two types of fermionic excitations have the same essential properties, and yield qualitatively similar pairing susceptibilities for triplets.

The triplet pairing instability can be captured by the effective action of the TI's electrons $\psi_{\tau\sigma}$:
\begin{eqnarray}\label{TriSeff}
S &=& \int \dd\widetilde{t} \, \dd^2 r \, \Biggl\lbrack
    \psi^\dagger \left( \frac{\partial}{\partial\widetilde{t}}
    + v\,\hat{{\bf z}}({\bf S}\times{\bf p})\tau^{z}+\Delta\tau^{x}-\mu \right) \psi \nonumber \\
&& + \sum_{\sigma} \left( V |\eta_\sigma^{\phantom{\dagger}}|^{2} + \eta_\sigma^{\phantom{\dagger}}
     \psi_{+\sigma}^{\dagger}\psi_{-\sigma}^{\dagger} + h.c. \right) \nonumber \\
&& + V'|\eta_0|^2 + \eta_0 \frac{\psi_{+\uparrow}^{\dagger}\psi_{-\downarrow}^{\dagger}
                                +\psi_{+\downarrow}^{\dagger}\psi_{-\uparrow}^{\dagger}}{\sqrt{2}} + h.c. \Biggr\rbrack \ .
\end{eqnarray}
We have eliminated all but the triplet interaction channels in favor of the renormalized couplings in this theory, and treat the electron dynamics as being relativistic at $\mu=0$ to a first approximation. The non-interacting spin-orbit Hamiltonian (\ref{SO}) in the first line has eigenstates $|{\bf p} \widetilde{\sigma} \widetilde{s}\rangle$ labeled by the following quantum numbers: momentum ${\bf p}$, ``helical'' spin $\widetilde{\sigma}$ (the eigenvalue of the operator $\widetilde{\sigma}^z = (\hat{\bf z}\times\hat{\bf p}) \boldsymbol{\sigma}$, where $\boldsymbol{\sigma}$ are spin Pauli matrices), and the band-index $\widetilde{s}=\pm 1$. The energy levels
\begin{equation}
E_{\widetilde{s}} ({\bf p}) = \widetilde{s}\epsilon({\bf p})-\mu \quad,\quad
\epsilon({\bf p}) = \sqrt{\frac{1}{4}|{\bf p}|^2v^2+\Delta^2}
\end{equation}
have no helical spin dependence owing to the existence of the orbital degree of freedom. The eigenstates can be obtained in a straight-forward fashion as superpositions of the pure orbital and spin $\sigma^z$ eigenstates $|{\bf p} \sigma \tau\rangle$. However, we need the inverted relationship in order to express the spin/orbital field operators $\psi_{\tau\sigma}$ in terms of the eigenstate field operators $\widetilde{\psi}_{\widetilde{s}\widetilde{\sigma}}$. We will derive it from the relationship between the $\sigma^z$ and $\widetilde{\sigma}^z$ eigenstates for a given momentum ${\bf p}$:
\begin{eqnarray}\label{Rel1}
|\widetilde{\sigma}\rangle &=& \frac{1}{\sqrt{2}} \left( e^{-i\frac{\varphi}{2}} |\uparrow\rangle +
  \widetilde{\sigma} e^{i\frac{\varphi}{2}} |\downarrow\rangle \right) \\
|\sigma\rangle &=& \frac{1}{\sqrt{2}} e^{i\sigma\frac{\varphi}{2}} \left( |\widetilde{\uparrow}\rangle +
  \sigma |\widetilde{\downarrow}\rangle \right) \nonumber \\[0.05in]
\varphi({\bf p}) &=& \textrm{arg}(-p_y+ip_x) \ , \nonumber
\end{eqnarray}
and the relationship between the pure orbital states $|\tau\rangle$ and band-eigenstates $|\widetilde{s}\rangle$ for any given ${\bf p}$ and $\widetilde{\sigma}$:
\begin{eqnarray}\label{Rel2}
\!\!\!\!\!\!\!\! |\widetilde{+}\rangle = A_{\widetilde{\sigma}}|+\rangle + B_{\widetilde{\sigma}}|-\rangle ~ &,& ~
  |\widetilde{-}\rangle = B_{\widetilde{\sigma}}|+\rangle - A_{\widetilde{\sigma}}|-\rangle \nonumber \\[0.04in]
\!\!\!\!\!\!\!\! |+\rangle = A_{\widetilde{\sigma}}|\widetilde{+}\rangle + B_{\widetilde{\sigma}}|\widetilde{-}\rangle ~ &,& ~
  |-\rangle = B_{\widetilde{\sigma}}|\widetilde{+}\rangle - A_{\widetilde{\sigma}}|\widetilde{-}\rangle
\end{eqnarray}
All states are normalized, and the coefficients $A_{\widetilde{\sigma}}$ and $B_{\widetilde{\sigma}}$ are:
\begin{equation}
A_{\widetilde{\sigma}} \!=\! \frac{\Delta}{\sqrt{2\epsilon(p)\lbrack \epsilon(p) + \widetilde{\sigma}\frac{pv}{2}\rbrack}}
  \quad , \quad
B_{\widetilde{\sigma}} \!=\! \frac{\epsilon(p) + \widetilde{\sigma}\frac{pv}{2}}
                                {\sqrt{2\epsilon(p)\lbrack \epsilon(p) + \widetilde{\sigma}\frac{pv}{2}\rbrack}}
\end{equation}
They have the following properties:
\begin{eqnarray}\label{Rel3}
A_{\widetilde{\sigma}}^2 + B_{\widetilde{\sigma}}^2 &=& 1 \\
  A_{\widetilde{\uparrow}}B_{\widetilde{\uparrow}} + \sigma\sigma' A_{\widetilde{\downarrow}}B_{\widetilde{\downarrow}}
    &=& \frac{\Delta}{\epsilon(p)} \delta_{\sigma\sigma'} \nonumber \\
A_{\widetilde{\uparrow}}^2 + \sigma\sigma' A_{\widetilde{\downarrow}}^2 &=& \delta_{\sigma\sigma'} -
      \frac{pv}{2\epsilon(p)} \delta_{\sigma,-\sigma'} \nonumber \\
B_{\widetilde{\uparrow}}^2 + \sigma\sigma' B_{\widetilde{\downarrow}}^2 &=& \delta_{\sigma\sigma'} +
      \frac{pv}{2\epsilon(p)} \delta_{\sigma,-\sigma'} \ . \nonumber
\end{eqnarray}

The Fourier transform of the bare electron Green's function $\langle T_{\widetilde{t}} \widetilde{\psi}_{\widetilde{s} \widetilde{\sigma}}^{\phantom{\dagger}}(0) \widetilde{\psi}_{\widetilde{s'} \widetilde{\sigma'}}^\dagger({\bf r},\widetilde{t}) \rangle$ constructed from the eigenstate field operators is:
\begin{equation}\label{BareGreen}
G_{\widetilde{s}}({\bf p},\omega) = \frac{1}{i\omega + \mu - \widetilde{s}\epsilon({\bf p})}
\end{equation}
in imaginary time ($\omega$ is the Matsubara frequency). Since the triplet Cooper pair fields $\eta_\sigma$ are coupled to the spin/orbital field operators $\psi_{\tau\sigma}$, we need the relationship between them and the eigenstate operators in order to obtain the vertex functions that appear in the pairing susceptibility Feynman diagram. Using (\ref{Rel1}) and (\ref{Rel2}) we find:
\begin{eqnarray}\label{Rel4}
\psi_{+\sigma} \!\!\! &=& \!\!\! \frac{1}{\sqrt{2}} e^{i\sigma\frac{\varphi}{2}} \! \left\lbrack
  A_{\widetilde{\uparrow}} \widetilde{\psi}_{\widetilde{+}\widetilde{\uparrow}} +
  B_{\widetilde{\uparrow}} \widetilde{\psi}_{\widetilde{-}\widetilde{\uparrow}} + \sigma \left(
  A_{\widetilde{\downarrow}} \widetilde{\psi}_{\widetilde{+}\widetilde{\downarrow}} +
  B_{\widetilde{\downarrow}} \widetilde{\psi}_{\widetilde{-}\widetilde{\downarrow}} \right) \right\rbrack \nonumber \\
\psi_{-\sigma} \!\!\! &=& \!\!\! \frac{1}{\sqrt{2}} e^{i\sigma\frac{\varphi}{2}} \! \left\lbrack
  B_{\widetilde{\uparrow}} \widetilde{\psi}_{\widetilde{+}\widetilde{\uparrow}} -
  A_{\widetilde{\uparrow}} \widetilde{\psi}_{\widetilde{-}\widetilde{\uparrow}} + \sigma \left(
  B_{\widetilde{\downarrow}} \widetilde{\psi}_{\widetilde{+}\widetilde{\downarrow}} -
  A_{\widetilde{\downarrow}} \widetilde{\psi}_{\widetilde{-}\widetilde{\downarrow}} \right) \right\rbrack \nonumber \\
\end{eqnarray}

\begin{figure}
\centering
\includegraphics[height=0.8in]{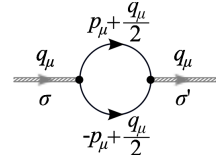}
\caption{\label{Bubble}The triplet-channel pairing Feynman diagram. The solid lines represent the electron propagators (\ref{BareGreen}). The spin-orbit coupling does not conserve spin, so the incoming $\sigma$ and outgoing $\sigma'$ Cooper pair spins can be different.}
\end{figure}

The inverse pairing susceptibility $\Pi_{\sigma\sigma'}(q_\mu)$ of inter-orbital triplets $\eta_\sigma$ is obtained from the Feynman diagram in the Fig.\ref{Bubble}:
\begin{eqnarray}
&& \!\!\!\!\! \Pi_{\sigma\sigma'}(q_\mu) = \int \!\! \frac{\dd^3 p_\mu}{(2\pi)^3} \Bigl\lbrack
    \langle T_{\widetilde{t}} \, \psi_{+\sigma}^{\phantom{\dagger}} \psi_{+\sigma'}^\dagger \rangle_{p_\mu^+}
    \langle T_{\widetilde{t}} \, \psi_{-\sigma}^{\phantom{\dagger}} \psi_{-\sigma'}^\dagger \rangle_{p_\mu^-} \nonumber \\
&& - \langle T_{\widetilde{t}} \, \psi_{+\sigma}^{\phantom{\dagger}} \psi_{-\sigma'}^\dagger \rangle_{p_\mu^+}
    \langle T_{\widetilde{t}} \, \psi_{-\sigma}^{\phantom{\dagger}} \psi_{+\sigma'}^\dagger \rangle_{p_\mu^-}
 \Bigr\rbrack + V\delta_{\sigma\sigma'}
\end{eqnarray}
where the noted field-operator contractions are calculated at the indicated internal momenta
\begin{equation}
p_\mu^\pm = \pm p_\mu + \frac{q_\mu}{2} \ ,
\end{equation}
and $p_\mu = ({\bf p},\omega)$, $q_\mu = ({\bf q},\Omega)$. Substituting (\ref{Rel4}) and using (\ref{Rel3}) we obtain:
\begin{eqnarray}\label{Pi2}
&& \!\!\!\!\!\!\! \Pi_{\sigma,\sigma}(q_\mu) = V + \frac{1}{2} \int\frac{\dd^3 p_\mu}{(2\pi)^3} \Biggl\lbrace \\
&& \!\!\!\!\! \left( \! 1 - \frac{\Delta^2}{\epsilon(p^-)\epsilon(p^+)} \! \right)\!\!
      \Bigl\lbrack G_{\widetilde{+}}^{\phantom{+}}(p_\mu^-) G_{\widetilde{+}}^{\phantom{+}}(p_\mu^+)
        + G_{\widetilde{-}}^{\phantom{+}}(p_\mu^-) G_{\widetilde{-}}^{\phantom{+}}(p_\mu^+) \Bigr\rbrack \!\! + \nonumber \\
&& \!\!\!\!\! \left( \! 1 + \frac{\Delta^2}{\epsilon(p^-)\epsilon(p^+)} \! \right)\!\!
      \Bigl\lbrack G_{\widetilde{-}}^{\phantom{+}}(p_\mu^-) G_{\widetilde{+}}^{\phantom{+}}(p_\mu^+)
        + G_{\widetilde{+}}^{\phantom{+}}(p_\mu^-) G_{\widetilde{-}}^{\phantom{+}}(p_\mu^+) \Bigr\rbrack \Biggr\rbrace \nonumber
\\[0.1in]
&& \!\!\!\!\!\!\! \Pi_{\sigma,-\sigma}(q_\mu) = \frac{e^{i\sigma\lbrack\varphi(p^-)+\varphi(p^+)\rbrack}}{2}
        \int\frac{\dd^3 p_\mu}{(2\pi)^3} \Biggl\lbrace \nonumber \\
&& \!\!\! - \frac{v^2 |{\bf p}^-| |{\bf p}^+|}{4 \epsilon(p^-)\epsilon(p^+)}
      \Bigl\lbrack G_{\widetilde{+}}^{\phantom{+}}(p_\mu^-) G_{\widetilde{+}}^{\phantom{+}}(p_\mu^+)
        + G_{\widetilde{-}}^{\phantom{+}}(p_\mu^-) G_{\widetilde{-}}^{\phantom{+}}(p_\mu^+) \Bigr\rbrack \nonumber \\
&& \!\!\! + \frac{v^2 |{\bf p}^-| |{\bf p}^+|}{4 \epsilon(p^-)\epsilon(p^+)}
      \Bigl\lbrack G_{\widetilde{-}}^{\phantom{+}}(p_\mu^-) G_{\widetilde{+}}^{\phantom{+}}(p_\mu^+)
        + G_{\widetilde{+}}^{\phantom{+}}(p_\mu^-) G_{\widetilde{-}}^{\phantom{+}}(p_\mu^+) \Bigr\rbrack \Biggr\rbrace
\nonumber
\end{eqnarray}
These expressions contain the Matsubara frequency integrals of the Green's function products:
\begin{eqnarray}
&& \!\!\!\!\! \int\frac{\dd \omega}{2\pi} \, \frac{1}{i\left(\frac{\Omega}{2}-\omega\right)+\mu-\widetilde{s}\epsilon(p^-)}
   \, \frac{1}{i\left(\frac{\Omega}{2}+\omega\right)+\mu-\widetilde{s}'\epsilon(p^+)} \nonumber \\
&& ~ = \frac{f\Bigl(\widetilde{s}\epsilon(p^-)-\mu\Bigr)-f\Bigl(-\widetilde{s}'\epsilon(p^+)+\mu\Bigr)}
          {-i\Omega+\widetilde{s}\epsilon(p^-)+\widetilde{s}'\epsilon(p^+)-2\mu} \nonumber \\
&& ~ \xrightarrow{T\to 0} -\frac{\delta_{\widetilde{s}\widetilde{s}'}}{\epsilon(p^-)+\epsilon(p^+)-\widetilde{s}(2\mu+i\Omega)} \ .
\end{eqnarray}
The Fermi distribution functions $f(\epsilon)=\lbrack 1+e^{\beta\epsilon} \rbrack^{-1}$ at zero temperature $T=\beta^{-1}\to 0$ yield a finite numerator in the result only when the two electron propagators carry energies of the same sign, $\widetilde{s}=\widetilde{s}'$ (note that the chemical potential $\mu$ lies inside the bandgap). This simplifies the calculations considerably, as only the first half of both expressions in (\ref{Pi2}) matters at $T=0$. In order to obtain $\Pi_{\sigma,-\sigma}$ we also need:
\begin{equation}
e^{i\lbrack\varphi(p^-)+\varphi(p^+)\rbrack} = \frac{e^{i2\varphi(q)}}{|{\bf p}^-| |{\bf p}^+|}
  \left\lbrack \left\vert\frac{\bf q}{2}\right\vert^2 - |{\bf p}|^2 e^{i2\theta} \right\rbrack \ ,
\end{equation}
where $\theta$ is the angle between the vectors ${\bf q}$ and ${\bf p}$. The term involving $\theta$ integrates out to zero in (\ref{Pi2}) because the rest of the expression is invariant under ${\bf p}\to-{\bf p}$. We finally perform analytic continuation $i\Omega \to \Omega+i0^+$ in order to eventually interpret the pole frequency $\Omega$ as mode energy, and obtain simplified formulas:
\begin{eqnarray}\label{Pi3}
\Pi_{\sigma,\sigma}(q_\mu) &=& V - \int \!\! \frac{\dd^2 p}{(2\pi)^2}
      \left( 1 - \frac{\Delta^2}{\epsilon({\bf p}^-)\epsilon({\bf p}^+)} \right) W({\bf p}^\pm) \nonumber \\[0.1in]
\Pi_{\sigma,-\sigma}(q_\mu) &=& e^{i2\sigma\varphi({\bf q})} \int \!\! \frac{\dd^2 p}{(2\pi)^2} \,
      \frac{|{\bf q}|^2v^2}{16\,\epsilon({\bf p}^-)\epsilon({\bf p}^+)} \, W({\bf p}^\pm) \nonumber \\[0.1in]
W({\bf p}^\pm) &=& \frac{\epsilon({\bf p}^-)+\epsilon({\bf p}^+)}
  {\lbrack\epsilon({\bf p}^-)+\epsilon({\bf p}^+)\rbrack^2-(2\mu+\Omega)^2}
  \nonumber \\[0.1in]
\varphi({\bf q}) &=& \textrm{arg}(-q_y+iq_x) \ ,
\end{eqnarray}
As before, ${\bf p}^\pm = \pm {\bf p} + \frac{1}{2}{\bf q}$, and the angle $\varphi({\bf q})$ corresponds to the direction of $\hat{\bf z}\times{\bf q}$. The above integral in $\Pi_{\sigma\sigma}(q_\mu)$ is ultra-violet divergent and has to be regularized. We may simply subtract from it the divergent part $\Pi_{\sigma\sigma}^{\textrm{reg}}$, which is calculated from the unregulated integral at ${\bf q}=0$, $\Omega=0$ by setting $\mu=\Delta=0$. This way we eliminate the contribution of unphysical vacuum fluctuations at arbitrarily short length-scales, an artifact of the continuum limit. We must keep $v$ finite, however, since it defines the electron spectrum at all energy scales in the present relativistic theory. This procedure yields
\begin{equation}
\Pi_{\sigma\sigma'}^{\textrm{reg}} = - \delta_{\sigma\sigma'} \int\frac{\dd^2 p}{(2\pi)^2} \frac{1}{pv} =
  -\frac{\Lambda}{2\pi v}\delta_{\sigma\sigma'} \ ,
\end{equation}
where $\Lambda$ is an explicitly introduced momentum cut-off.

By the same approach we can obtain the inverse pairing susceptibility in the symmetric triplet channel $\eta_0$:
\begin{equation}\label{Pi3b}
\Pi_{0,0}(q_\mu) = V' - \int \!\! \frac{\dd^2 p}{(2\pi)^2}
  \left( 1 - \frac{\frac{v^2}{4} |{\bf p}^-| |{\bf p}^+| + \Delta^2}
     {\epsilon({\bf p}^-)\epsilon({\bf p}^+)} \right) W({\bf p}^\pm)
\end{equation}
which requires no regularization. All quadratic couplings between $\eta_\pm$ and $\eta_0$ are found to be zero.

The imaginary parts of the expressions (\ref{Pi3}) and (\ref{Pi3b}) can be calculated analytically by adding an infinitesimal imaginary part $i0^+$ to the real frequency $\Omega$. The real parts of (\ref{Pi3}) and (\ref{Pi3b}) can then be deduced in principle by applying the Kramers-Kronig relations, but this is severely complicated by the presence of terms in $\textrm{Im} \lbrace \Pi_{\sigma\sigma'} \rbrace$ that do not vanish at $|\Omega|\to\infty$. We instead carry out all subsequent analysis by the simpler numerical evaluation.

The relativistic nature of the Cooper pair spectrum is already evident from (\ref{Pi3}). There are Cooper pairs of particles and pairs of holes, which are energetically equivalent when the electron spectrum has particle-hole symmetry ($\mu=0$). Also, the factor $e^{i2\sigma\varphi({\bf q})}$ in $\Pi_{\sigma,-\sigma}(q_\mu)$ reveals that the triplet pairs are spin-orbit-coupled and have a twice larger spin-orbit SU(2) charge than electrons. The effective action for triplets takes the form:
\begin{equation}\label{EffAct}
S_{\eta} = \int \frac{\dd^3 q_\mu}{(2\pi)^3} \Bigl\lbrack \eta^\dagger \Pi \, \eta + \cdots \Bigr\rbrack \ ,
\end{equation}
where $\eta = ( \eta_+, \eta_0, \eta_- )$ is the spinor of triplet Cooper pair fields and
\begin{equation}\label{PiMatrix}
\Pi =
  \left( \begin{array}{ccc}
    \Pi_{++} & 0 & \Pi_{-+} \\
    0 & \Pi_{00} & 0 \\
    \Pi_{+-} & 0 & \Pi_{--}
  \end{array} \right) \ .
\end{equation}
Note that $\Pi_{++}(q_\mu) = \Pi_{--}(q_\mu)$ and $\Pi_{-+}^{\phantom{*}}(q_\mu) = \Pi_{+-}^*(q_\mu)$. Diagonalizing this matrix reveals three modes: the helical triplets $\widetilde{\eta}_\pm$, whose quadratic coupling is:
\begin{equation}\label{Pi4}
\widetilde{\Pi}_\pm(q_\mu) = \Pi_{\sigma,\sigma}(q_\mu) \pm |\Pi_{\sigma,-\sigma}(q_\mu)| \ ,
\end{equation}
and the symmetric triplet $\eta_0$ whose quadratic coupling is (\ref{Pi3b}). Expanding $\Pi_{\sigma\sigma'}$ and $\Pi_{00}$ to the quadratic order in the Cooper pair's frequency and momentum reveals:
\begin{equation}\label{PiMatrix2}
\Pi = A - B(\Omega+2\mu)^2 + c \Bigl\lbrack\hat{{\bf z}}({\bf q}\times{\bf S})\Bigr\rbrack^{2} + Dq^2 + \cdots \ ,
\end{equation}
where lowercase $a,b,c\dots$ denote real constants, and uppercase $A = a' + a''\Phi_0^2$, $B = b'+b''\Phi_0^2$, $D = d'+d''\Phi_0^2$ denote matrices involving the Yang-Mills magnetic flux (\ref{Flux}). The spin operator $\bf S$ components in the above equation are given by (\ref{SpinOperators}). Therefore, our perturbative calculation reveals the Rashba spin-orbit coupling for triplets. The dispersions of the spin-orbit coupled particle and antiparticle modes are merged into
\begin{equation}
\Pi \propto \Bigl\lbrack \Omega+2\mu-v'\hat{{\bf z}}({\bf q}\times{\bf S}) \Bigr\rbrack
  \Bigl\lbrack -\Omega-2\mu-v'\hat{{\bf z}}({\bf q}\times{\bf S}) \Bigr\rbrack + \cdots \nonumber
\end{equation}
If we wished to extract the dynamics of only the particle or only the hole modes, the ensuing action would qualitatively look like (\ref{LG2}). Note that our present calculation does not feature the SU(2) gauge symmetry. We broke it explicitly in (\ref{TriSeff}) to simplify calculations, so independent terms such as $D$ are allowed in (\ref{PiMatrix2}).

Let us now analyze the triplet spectrum. The absence of a solution to $\widetilde{\Pi}_\pm(q_\mu)=0$ in (\ref{Pi4}) would indicate that a coherent helical bosonic mode does not exist, while $\widetilde{\Pi}_\pm(q_\mu)<0$ at $\Omega=0$ indicates pairing instability. We will numerically analyze the mode dispersions by expressing $\widetilde{\Pi}_\pm(q_\mu)$ in the scaling form:
\begin{equation}\label{Pi5}
\widetilde{\Pi}_\pm(q_\mu) = V F_\pm \left(\frac{\Omega+2\mu}{Vv^2},\frac{q}{Vv};\frac{\Delta}{Vv^2}\right) \ .
\end{equation}
The interaction parameter $V$ has the dimensions of mass in $d=2$, and is normally positive and inversely proportional to the strength of triplet-channel attractive interactions induced by the proximity effect. By applying a gate voltage, $V$ can be turned negative. Recall that the functional form (\ref{Pi5}) that we obtain from the one-loop approximation is accurate to all orders of perturbation theory when expressed in terms of the renormalized parameters (because the ground state is fully gapped).

\begin{figure}
\centering
\subfigure[{}]{\includegraphics[height=2.7in]{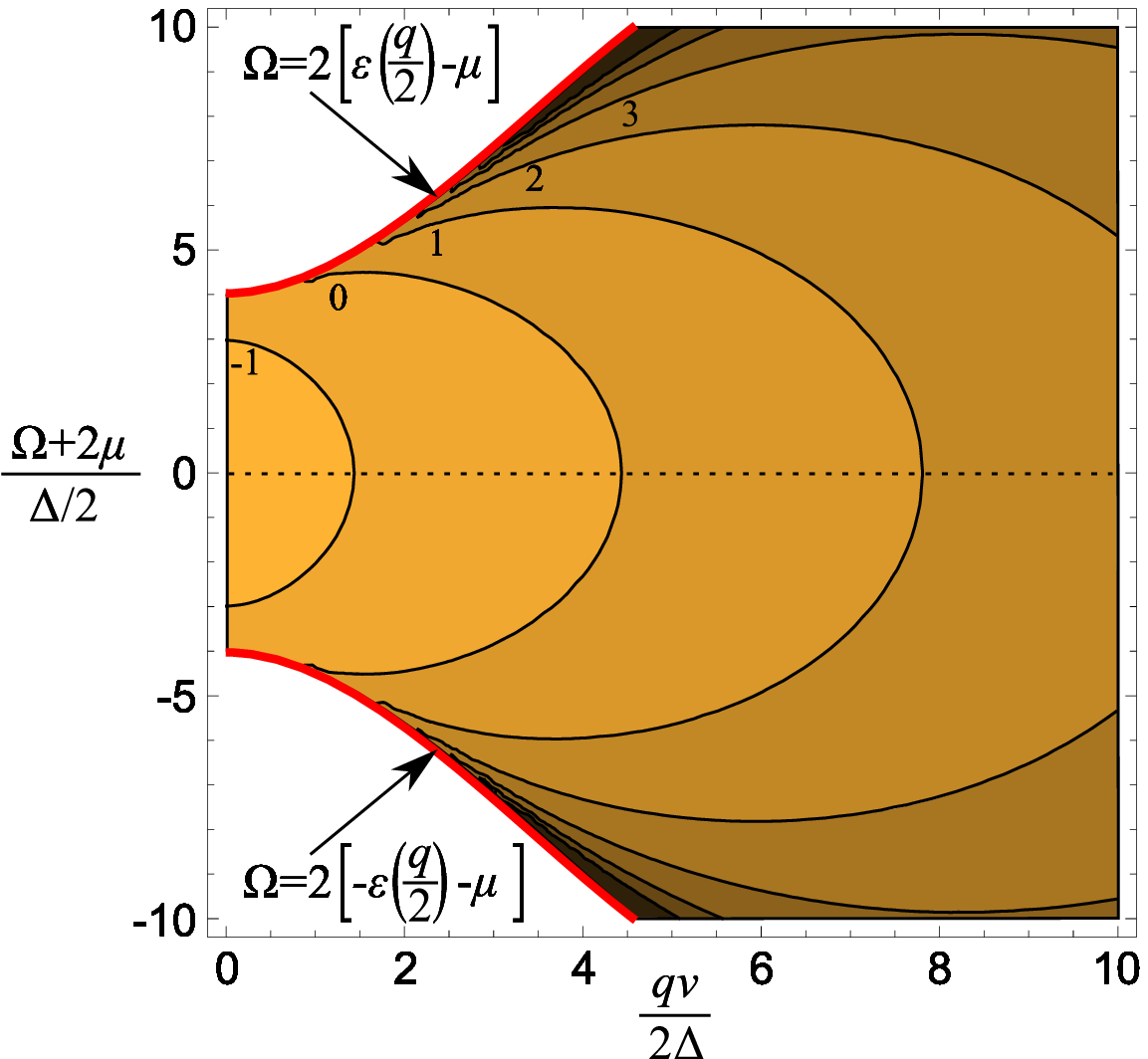}}
\hspace{0.2in}
\subfigure[{}]{\includegraphics[height=2.7in]{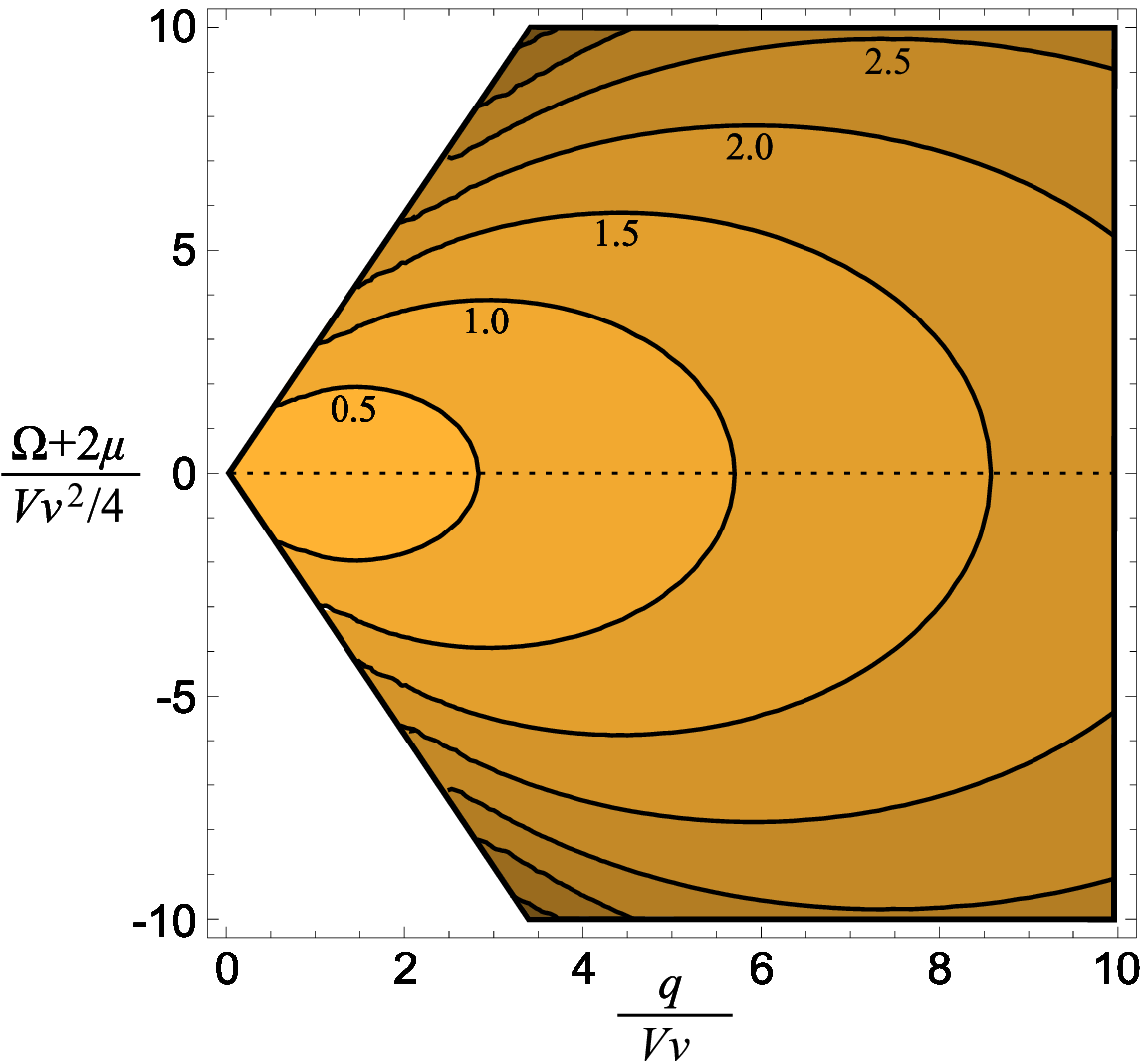}}
\caption{\label{edisp2D}Energy dispersions $\Omega_-(q)$ of the critical helical triplet mode, parametrized by the rescaled inverse interaction strength $Vv^2/2\Delta$ at a fixed bandgap $\Delta$ in (a), and by $4\Delta/Vv^2$ at a fixed $V$ in (b). Unshaded regions indicate the continuum of unpaired electron states where $\textrm{Im}(\widetilde{\Pi}_-)\neq 0$. Increasing $q$ and following a contour eventually brings a finite mode energy to zero at some $q=q_{\textrm{min}}$ where condensation can occur. All modes beyond $q>q_{\textrm{min}}$ are condensed as well because $\widetilde{\Pi}_-(q,0)$ is negative. This massive instability shows that the triplets tend to condense at finite momenta, where their modes are rotationally degenerate. Only if $q_{\textrm{min}} > \Lambda$, the momentum cut-off, the modes that would condense do not exist and the insulating state remains a band-insulator. Note that $q_{\textrm{min}}$ grows by increasing the bandgap or decreasing the interaction strength, and that all contours in the pane (b) can be collapsed to one by geometric scaling.}
\end{figure}

\begin{figure}
\centering
\includegraphics[height=2.7in]{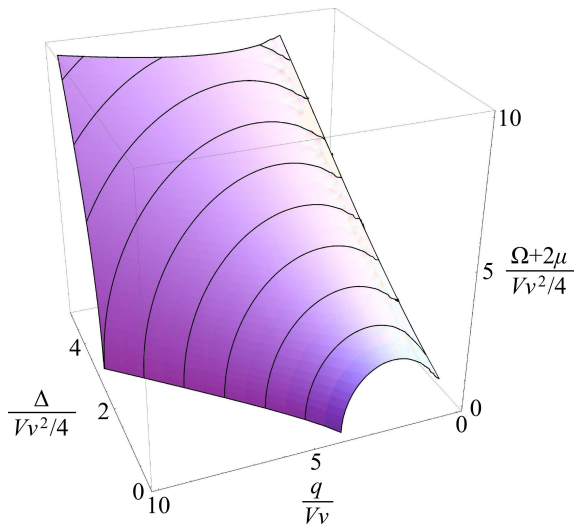}
\caption{\label{edisp3D}The critical triplet mode dispersion $\Omega_-(q)$ for arbitrary chemical potential $\mu$, quantum well bandgap $\Delta$, and inverse interaction strength $V$. This is just the $\widetilde{\Pi}_-(q,\Omega)=0$ surface obtained from (\ref{Pi5}), or the three-dimensional version of the contour plot in Fig.\ref{edisp2D}(b) for $\Omega+2\mu>0$. The curved lines are dispersions at fixed $\Delta$.}
\end{figure}

We will not make reference to particle versus hole character of the Cooper pairs in the following discussion, but rather exploit the symmetry under $\Omega+2\mu \to -(\Omega+2\mu)$. If both helical modes $\widetilde{\eta}_\pm$ were coherent, they would have the same and positive energy at zero momentum, but the energy of $\widetilde{\eta}_+$ would increase, while the energy of $\widetilde{\eta}_-$ would decrease with momentum. The zero-momentum energy gap is often high enough to lie in the unpaired electron continuum, and then the $\widetilde{\eta}_+$ is never coherent while $\widetilde{\eta}_-$ becomes coherent only at sufficiently large momenta. But generally, even if both modes exist, only $\widetilde{\eta}_-$ is a candidate for condensation, and generally only at sufficiently large momenta where its energy approaches zero. The Figures \ref{edisp2D} and \ref{edisp3D} show the $\widetilde{\eta}_-$ energy dispersion $\Omega_-(q)$ obtained from $\widetilde{\Pi}_-(q,\Omega)=0$. The smallest momentum at which condensation occurs is $q_{\textrm{min}}$, where $\Omega_-(q_{\textrm{min}})=0$. However, the modes at larger momenta are also unstable, because $\widetilde{\Pi}_-(q,0)<0$ at $q>q_{\textrm{min}}$. This means that the assumed band-insulating ground-state in calculations gives way to a non-uniform superconductor in which the helical triplet Cooper pairs condense at finite momenta.

Interactions between Cooper pairs must resolve the massive degeneracy of simultaneously condensing modes with different momentum orientations. These modes have even larger density of states if the condensation occurs at $\mu=0$, which can be seen in Fig.\ref{edisp2D} by the vertical tangents of $\Omega_-(q)$ at $\Omega=0$, the courtesy of relativistic particle-hole symmetry. But, most generally, the presence of a large number of unstable modes at $q>q_{\textrm{min}}$ means that Cooper pair interactions must play an important role in stabilizing the ground-state, regardless of the fact that in real materials the crystalline lattice reduces the rotational mode degeneracy to a finite discrete symmetry group at large momenta.

\begin{figure}
\subfigure[{}]{\includegraphics[height=2.7in]{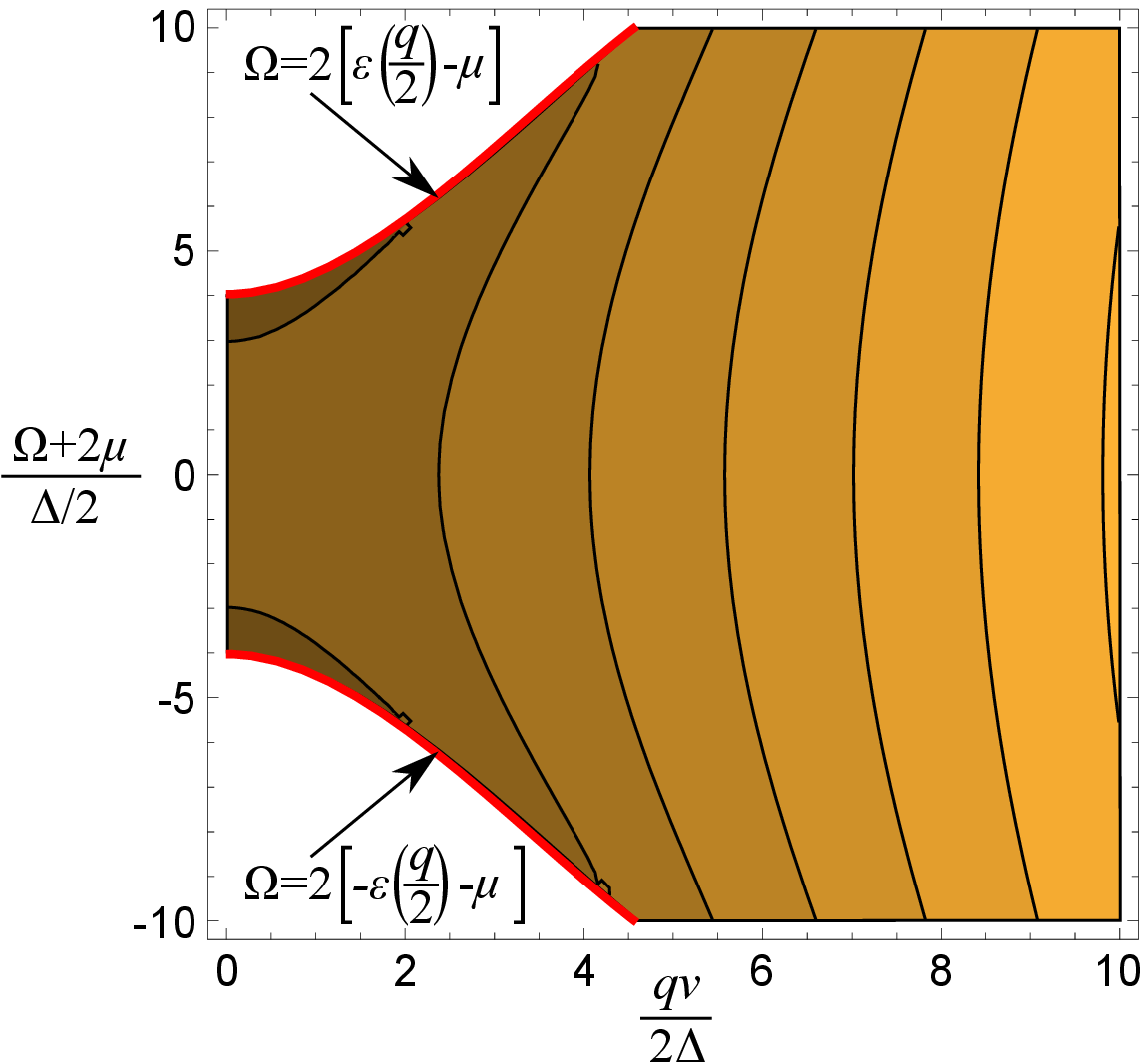}}
\subfigure[{}]{\includegraphics[height=2.7in]{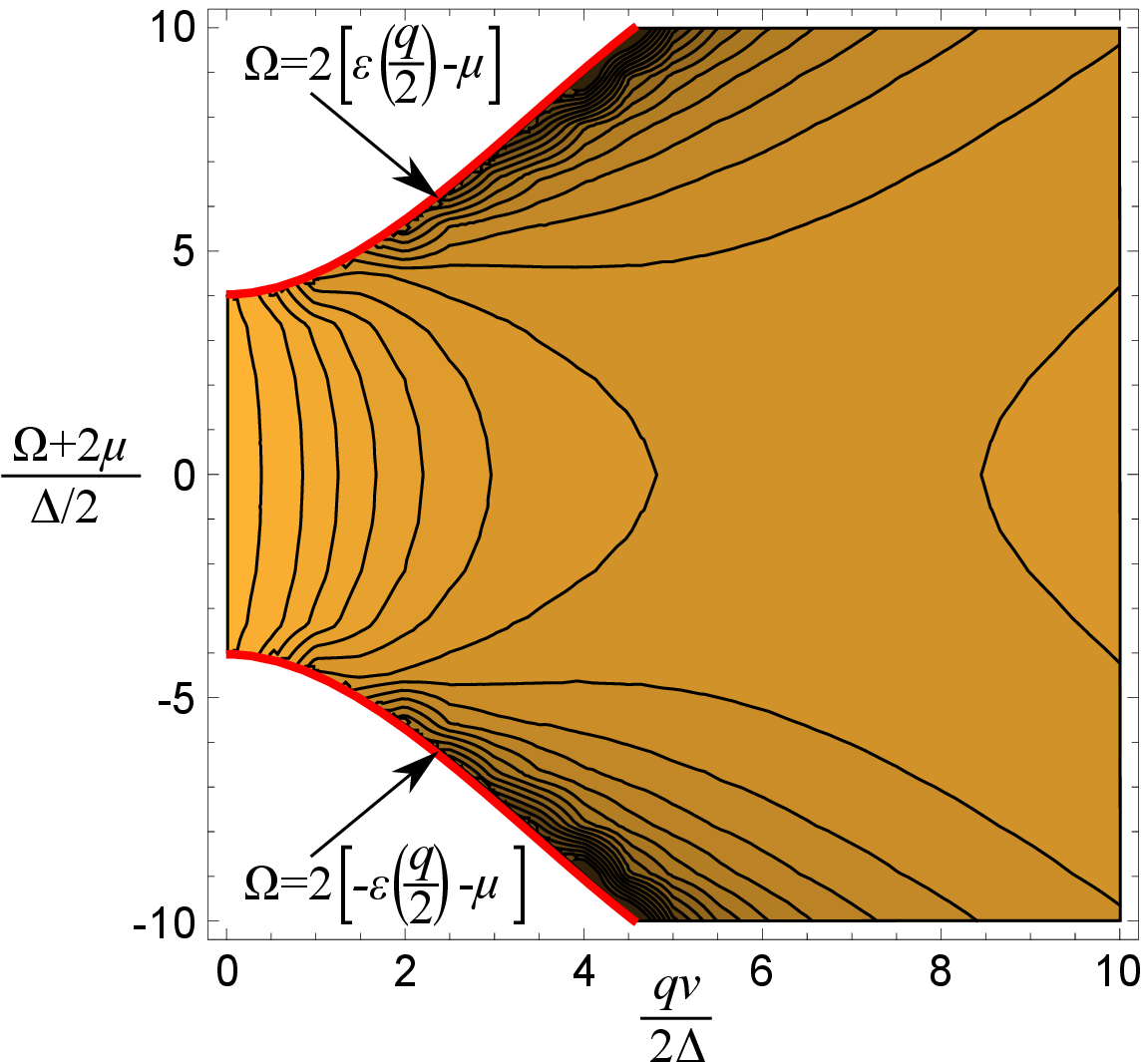}}
\caption{\label{edispHiModes}Energy dispersions $\Omega_+(q)$ in (a) and $\Omega_0(q)$ in (b) of the high-energy helical triplet mode $\widetilde{\eta}_+$ and the symmetric triplet mode $\eta_0$ respectively. The TI's bandgap $\Delta$ is fixed, while the contours and brightness correspond to the interaction strengths, as in the Fig.\ref{edisp2D}(a).}
\end{figure}

For completeness, we show the dispersions of the high-energy helical triplet $\widetilde{\eta}_+$ and symmetric triplet mode $\eta_0$ in the Figure \ref{edispHiModes}. These modes need not exist as coherent excitations, especially $\widetilde{\eta}_+$. However, it is useful to compare its calculated dispersion (as if it were coherent) with that of $\widetilde{\eta}_-$. The energy of $\widetilde{\eta}_+$ grows with momentum as is usual, while the energy of $\widetilde{\eta}_-$ decreases with momentum. The symmetric triplet $\eta_0$ has a tendency to condense at finite momenta just like $\widetilde{\eta}_-$.

\section{Phase diagram}\label{secPhDiag}

Here we combine the information gathered so far to outline the qualitative features of the phase diagram. Our main interest are the phases of the helical triplet Cooper pair mode $\widetilde{\eta}_-$, which is created in the TI quantum well by the proximity effect and then energetically empowered by the Rashba spin-orbit coupling. If an $\widetilde{\eta}_-$ mode carries momentum $\bf q$, then it also carries spin in the direction $\hat{\bf z}\times{\bf q}$.

Let us first briefly explain why the finite momentum condensation of helical triplets likely leads to the formation of a vortex lattice. If the helical condensate formed at only one particular momentum ${\bf q}$, it would carry uniform charge and spin currents. This is not possible in equilibrium. In order to eliminate the charge current, an equally strong condensation should also occur at the momentum $-{\bf q}$. The ensuing condensate would be a Fulde-Ferrell-Larkin-Ovchinnikov (FFLO) state. However, the condensed helical triplets have their spin locked to momentum by the right-hand rule, so a finite uniform spin supercurrent survives in this state and makes it unacceptable in equilibrium. In fact, after the charge flow is removed, the spin flow cannot be eliminated due to the spin-momentum locking as long as $\widetilde{\eta}_-$ is the only condensed spinful mode at finite momenta. Therefore, the equilibrium helical triplet condensate must organize its spin superflow into closed loops.

Vorticity is quantized in any superconducting state. The unavoidable quadratic kinetic energy $E_{\textrm{kin}}(q) = K_{\textrm{t}} q^2$ (which was temporarily removed from (\ref{TriSeff}) in order to simplify pairing calculations) always makes the single-quantized vortices least costly. Namely, the vortex core energy is logarithmically ultra-violet divergent in the vortex core radius, and proportional to the vortex ``charge'' squared. Therefore, the equilibrium state of triplet spin-currents has the minimum energy in some Abrikosov lattice of single-quantized vortices. It should be also noted that the SU(2) flux due to the Rashba spin-orbit coupling cannot be expelled from the system by the presence of supercurrents (our gauge field is non-dynamical, there are no gauge bosons analogous to photons). Consequently, the triplet superconductor in the TI is of type-II. Having a spin-current flow only along the system boundary would yield too small energy gain, proportional only to the system's perimeter instead of its area.

We may regard these vortices as SU(2) topological defects if the TR-symmetry is respected \cite{Nikolic2011a}. This TR-invariant state of $\widetilde{\eta}_-$ is such a superposition of $S^z$ eigenmodes $\eta_+$ and $\eta_-$ that each U(1) vortex defect in the phase of $\eta_+$ coincides with an U(1) anti-vortex in the phase of $\eta_-$. Another possible outcome of triplet condensation in equilibrium is a vortex lattice of \emph{charge} supercurrents with a spin texture locally perpendicular to the current flow. This state would spontaneously break the TR symmetry. Analyzing the structure and energetics of such vortex lattices goes beyond the scope of this paper and will be pursued elsewhere. We will only note here that related vortex lattice states have been found in two-component boson model calculations \cite{Cole2012, Radic2012}.

The perturbative picture of the triplet pairing instability (pertaining to the case of absent singlet condensate) is fully self-consistent only if the insulating state adjacent to the transition is a band-insulator, as we initially assumed. The band-insulator then results from the smallest condensation momentum $q_{\textrm{min}}$ being larger than the momentum cut-off $\Lambda$ set by the material's crystal lattice. All existing helical modes at $q<\Lambda$ are then gapped. However, the above expectation of a vortex lattice in the superconducting phase carries the seed of the perturbation theory's breakdown near the transition. Namely, quantum fluctuations are capable of melting the vortex lattice in a first-order transition, which preempts the naively obtained second order pairing transition \cite{Nikolic2011a}. This is a fundamentally non-perturbative effect missed by our calculation, as it involves the quantum fluctuations of topological defects. Cooper pairs remain dynamically dominant at low energies in the ensuing vortex liquid state, and hence we may view the obtained insulator as strongly correlated. It has been argued that such insulators could be exotic incompressible quantum liquids with fractional excitations. Similar scenarios have been numerically observed for bosonic particles in effective magnetic fields: quantum melting of a vortex lattice can produce fractional quantum Hall states \cite{Wilkin2000, Cooper2001, Regnault2003, Chang2005, Cooper2008}.

\begin{figure}
\centering
\includegraphics[height=2.9in,width=2.9in]{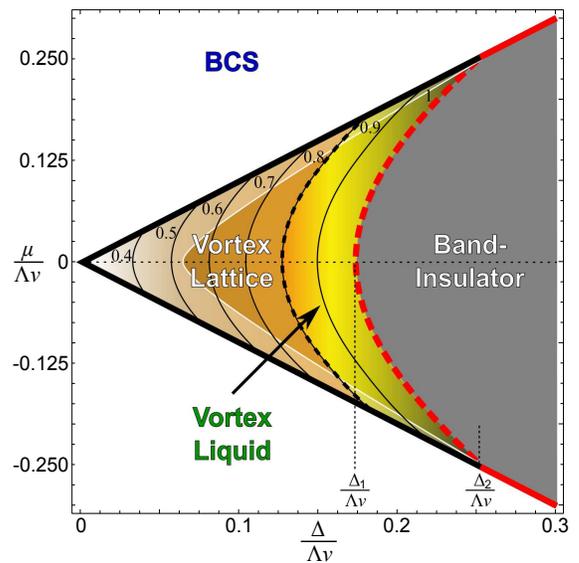}
\caption{\label{PhDiag2}The phase diagram of the helical triplet superconductivity in the TI at a fixed interaction strength $V$. The unshaded areas are metallic in the absence of interactions, and become superconducting with any amount of attractive interactions via the BCS instability. The shaded colored area is a helical superconductor with an SU(2) vortex lattice, which can coexist with a singlet superconductor shaded in the lighter color as discussed in the section \ref{secCompetition}. The relative positions of singlet and triplet transitions cannot be determined in the present qualitative calculation. Quantum fluctuations can melt the SU(2) vortex lattice and yield a strongly correlated topological insulator of Cooper pairs. The labeled contours correspond to fixed values of $q_{\textrm{min}}/\Lambda$ calculated from (\ref{Pi4}), where $\Lambda$ is the momentum cut-off. When $q_{\textrm{min}}>\Lambda$, the ground-state is a band-insulator shaded in gray.}
\end{figure}

It now remains to explore how the helical triplet superconductivity could be brought into competition with correlated insulating states. The Figure \ref{PhDiag2} shows the pairing instability phase diagram in terms of the two practical tuning parameters in the device from Fig.\ref{Device}, the quantum well bandgap $\Delta$ (controlled by the well thickness), and the chemical potential $\mu$ (tuned by the gate voltage). The present model features two special bandgap values $\Delta_{1,2}$ at any fixed interaction strength and momentum cut-off. If $\Delta<\Delta_1$, the quantum well superconducts for any applied gate voltage, and if $\Delta>\Delta_2$ the ground-state is either a band-insulator or a BCS superconductor depending on $\mu$. Keeping $\Delta<\Delta_2$ and placing $\mu$ within the colored area in Fig.\ref{PhDiag2} yields a vortex lattice superconducting state. The perturbation theory views the phase transition between such a superconductor and the band-insulator as a second-order one (shown as the dashed thick red line): coming from the insulator side ($\Delta<\Delta_2$), a helical triplet mode is coherent and its gap gradually closes at $q=\Lambda$. However, this picture is not entirely accurate as we explained above. Coming from the superconductor side, the quantum zero-point motion of vortices leads to a first-order vortex lattice melting transition (at some critical finite value of the order parameter, shown by the thin dashed black line) before the perturbative second-order transition can take place. This means that the there is at least one intervening correlated insulator phase between the superconductor and the band-insulator, a quantum vortex liquid which could be a fractional topological insulator.

\section{Discussion and conclusions}\label{secDiscussion}

In conclusion, we showed that unconventional triplet superconductivity is generally possible in TI quantum wells proximate to a conventional superconductor. It involves triplet condensation at finite momenta under the influence of the Rashba spin-orbit coupling, likely leading to a vortex lattice state. Quantum melting of such a vortex lattice is tunable by the gate voltage and expected to yield a correlated triplet insulator, a vortex liquid that could exhibit topological order and excitations with fractional statistics. We argued that singlet superconductivity, also possible in the TI quantum well, need not be an obstacle to obtaining the superconducting and correlated insulating phases of triplets.

Since the correlated insulators necessarily intervene between the triplet superconductor and band-insulator phases, there are no fundamental obstacles to realizing them in practical devices. Apart from disorder, the main practical limitations are how large bandgaps and how strong proximity-induced interactions can be engineered. If the proximity effect can produce superconductivity in the TI in the first place, which is the current premise made in literature, then the interactions are strong enough as well. One must then engineer a sufficiently thin quantum well, with a sufficiently large bandgap, to allow approaching the band-insulator in the Fig.\ref{PhDiag2}. The required bandgap may be larger than a single TI monolayer could provide. If so, one must weaken the proximity-induced interactions in the TI, and this is guarantied to expand the band-insulator region in the phase diagram. Controlling interactions may not be easy, but making them weaker is surely possible by choosing a SC material with a lower critical temperature, inserting tunneling barriers between the SC and TI, etc.

We will finish this analysis by discussing the role of spinless intra-band and inter-band Cooper channels in the TI quantum well. The idealized model (\ref{SO}) has a U(1) symmetry associated with the conservation of the helical spin projection $\widetilde{\sigma}^z$. This gives the spinful helical Cooper pairs $\widetilde{\eta}_\pm$ a certain autonomy, thereby justifying our focus on them in the section \ref{secTriplet}. We should, however, ask how the spinful and spinless channels might compete with each other. In the absence of a full spin SU(2) symmetry, the spinless channels can be coupled with one another. Among their mixtures, the modes with pronounced inter-orbital singlet $\phi_0$ or symmetric triplet $\eta_0$ components are unlikely to be dynamically important anywhere in the phase diagram. This is because the lowest energy $\phi_0$ and $\eta_0$ Cooper pairs of not too large size are made from two electrons with roughly opposite momenta, opposite spins and opposite orbitals, which tends to place them in the different spin-orbit bands and impose the full bandgap as a hurdle to pairing by weak interactions. The intra-band singlets are the main competitors to helical triplets, but they are not aided by the spin-orbit coupling.

Realistic TIs do not have the spin U(1) symmetry, at least due to disorder. For this reason, a realistic model should not conserve the number of any type of Cooper pairs (only their total charge is conserved), and terms such as $\eta_+^\dagger \phi_0^{\phantom{\dagger}}$ are allowed in the action (\ref{LG2}). Consequently, a condensate driven by any channel will have phase-locked contributions from other channels. There is a certain frustration to be resolved, because the spinless Cooper pairs prefer to condense into a uniform state, while the spinful ones prefer to form a vortex lattice. Indeed, depending on which condensate component is dominant, the superconducting state will either be uniform or host a vortex lattice, and its transformation under translations, rotations, etc. is the only qualitative property that can distinguish it from the other superconducting states. However, the phase-locking is not efficient in the TR-invariant vortex lattice state dominated by the spinful triplets. Namely, this state features finite but opposite circulating charge supercurrents of $\eta_+$ and $\eta_-$ that cancel each other's drag of supercurrents in the spinless channels. Only if the TR symmetry were spontaneously broken would there be some net drag, but the additional energy required to support circulating supercurrents in the channels where they are unnatural likely makes the TR-invariant vortex lattice (of spinful triplets alone) the most favorable ground-state.

\section{Acknowledgments}

We are very grateful to Michael Levin for insightful discussions, and to the Aspen Center for Physics for its hospitality. This research was supported by the Office of Naval Research (grant N00014-09-1-1025A), the National Institute of Standards and Technology (grant 70NANB7H6138, Am 001), and the U.S. Department of Energy, Office of Basic Energy Sciences, Division of Materials Sciences and Engineering under Award DE-FG02-08ER46544.




\end{document}